\pdfoutput=1
\documentclass[aps,pre,superscriptaddress,amsmath,amssymb,reprint,nofootinbib,twoside]{revtex4-2}
\usepackage{graphicx}
\usepackage{bm}
\newcommand{\circledbigcirc}{\mbox{$\bigcirc$\hspace{-0.9em}\raise0.2ex\hbox{$\scriptstyle{\bigcirc}$}\hspace{0.1em}}}

\begin{document}

\title{Order-chaos transition in correlation diagrams and quantization of period orbits}
\author{F. J. Arranz}
\email{fj.arranz@upm.es (corresponding author)}
\affiliation{Grupo de Sistemas Complejos, Universidad Polit\'ecnica de Madrid, av.\ Puerta de Hierro 2--4, 28040 Madrid, Spain}
\author{J. Montes}
\email{jmontes.3@alumni.unav.es}
\affiliation{Grupo de Sistemas Complejos, Universidad Polit\'ecnica de Madrid, av.\ Puerta de Hierro 2--4, 28040 Madrid, Spain}
\affiliation{Departamento de Qu\'\i mica, Universidad Aut\'onoma de Madrid, Cantoblanco, 28049 Madrid, Spain}
\author{F. Borondo}
\email{f.borondo@uam.es}
\affiliation{Departamento de Qu\'\i mica, Universidad Aut\'onoma de Madrid, Cantoblanco, 28049 Madrid, Spain}
\date{\today}

\begin{abstract}
Eigenlevel correlation diagrams has proven to be a very useful tool to understand eigenstate characteristics of classically chaotic systems. In particular, we showed in a previous publication [Phys.\ Rev.\ Lett.\ {\bf 80}, 944 (1998)] how to unveil the scarring mechanism, a cornerstone in the theory of quantum chaos, using the Planck constant as the correlation parameter. By increasing Planck constant, we induced a transition from order to chaos, in which scarred wavefunctions appeared as the interaction of pairs of eigenstates in broad avoided crossings, forming a well defined frontier in the correlation diagram. In this paper, we demonstrate that this frontier can be obtained by means of the semiclassical quantization of the involved scarring periodic orbits. Additionally, in order to calculate the Maslov index of each scarring periodic orbit, which is necessary for the semiclassical quantization procedure, we introduce a novel straightforward method based on Lagrangian descriptors. We illustrate the theory using the vibrational eigenstates of the LiCN molecular system.
\end{abstract}

\maketitle

\section{\label{sec:intro}Introduction}

A cornerstones in the development of quantum chaos is level statistics~\cite{Haake.quantum.chaos.book}, i.e., the statistics of the spacing between adjacent energy levels in the spectrum of a Hamiltonian operator, which is mathematically based on the random matrix theory~\cite{random.matrix.handbook}. In this framework, two extreme cases can be considered: (i) integrable systems where, as was shown by Berry and Tabor~\cite{BT.conjecture}, the statistics of the energy spacing follows a exponential distribution derived from a Poisson distribution of the eigenenergies, and (ii) classically fully chaotic systems (K-systems) where, as was conjectured by Bohigas, Giannoni, and Schmit~\cite{BGS.conjecture} and proved in a semiclassical context by M\"uller et al.~\cite{BGS.conjecture.proof2}, the statistics of the energy spacing follows the distribution of the Gaussian orthogonal ensemble (GOE) of the random matrix theory. In the first case, since the distribution of levels corresponds to a Poisson process, there are no interaction between energy levels, such that small spacing is highly likely. While, in the second case, the GOE distribution implies certain {\em level repulsion}, such that small spacing is highly unlikely.

However, in most cases, physical systems are not fully chaotic (K-systems), but systems with mixed dynamics, i.e., systems where chaotic regions and islands of regularity coexist in classical phase space. This case is an intermediate one where it can typically be found an order region, with a Poisson-like statistics, and a mixed chaos region, with an in-between Poisson-GOE statistics that can be described by means of a Weibull distribution, as was proposed by Brody~\cite{Brody.distribution} (therefore known as the Brody distribution in this context), which smoothly interpolates between both extreme cases. In the mixed chaos region, there are states with very little or no interaction, which are related to classical islands of stability, and also states with level repulsion, which are related to classical chaotic regions. Molecular systems, in particular, mostly belong to this intermediate case.

Moreover, notice that by taking the Planck constant as a varying parameter, thus obtaining a correlation diagram of eigenenergies versus Planck constant, it can be used as an ideal tool to implement a kind of microscope that focuses with varying resolution on the classical regular structures existing in the phase space of systems with mixed chaos. In this way, it has been shown in the literature~\cite{Arranz.LiCN.hbar.correlation1,Arranz.LiCN.hbar.correlation2,Arranz.LiCN.HCN.HO2.hbar.correlation,Parraga.KCN.hbar.correlation} for different molecular systems, including HCN, LiCN, KCN, and HO$_2$, the existence of a singular series of broad avoided crossings (ACs) in the correlation diagram of eigenenergies versus Planck constant, which constitutes the frontier that separates the order and mixed chaos regions. Namely, below this series of ACs (the order region), no level repulsion is found, while above the series of ACs (the mixed chaos region), extensive level repulsion is found, as well as some states with very little interaction related to classical islands of stability. Then, based on the level statistics results, this series of ACs can be considered the frontier between order and chaos. Interestingly, the eigenstates involved in this frontier are scarred states, i.e., states where the extremes (maxima and minima) of their wavefunctions are distributed along an {\em isolated unstable} periodic orbit (PO), phenomenon first studied by Heller~\cite{Heller.scars} in the Bunimovich stadium billiard. Notice that, as energy increases, the eigenstates of the frontier are the first scarred states to appear, such that this case adds some insight into the scar formation in molecular systems.

In a previous work~\cite{Arranz.LiCN.correspondence.resonances}, we studied the correspondence between classical and quantum resonances in the order region of the correlation diagram of the LiCN molecular system, leading to a semiclassical theory for this correspondence. In the present work, we further study the frontier of scars %
\pagebreak
quantitatively in the correlation diagram of the LiCN molecular system, addressing the semiclassical quantization of the scarring POs involved in this frontier of scars, which will lead to obtaining a semiclassical frontier between order and chaos.

The semiclassical quantization of POs was addressed by Gutzwiller~\cite{Gutzwiller.trace.formula} in the seminal work where his celebrated trace formula was first obtained. The quantization condition obtained for the classical action depends on the number of conjugate points of the PO over one period, an integer value also known as Maslov index. Therefore, in order to achieve the semiclassical quantization of a PO, it is necessary to compute its Maslov index, but a rigorous calculation using any of the different methods described in the literature is mathematical complicated. In the present work, as an additional result, we will introduce a novel straightforward method for the calculation of the Maslov index of a PO, based on Lagrangian descriptors~\cite{Mancho.LDs.different_definitions,Mancho.LDs.theoretical_proofs}. Lagrangian descriptors have been shown to be a fruitful tool to study the complex invariant structures existing in the phase space of non-linear systems. In recent years, these mathematical objects have been applied to a multitude of cases, including the LiCN molecular system~\cite{Revuelta.LDs.LiCN.1,Revuelta.LDs.LiCN.2}. Now, we introduce this novel application of Lagrangian descriptors, while a detailed research will be presented in a further work.

The organization of the paper is as follows. Section~\ref{sec:system.calculations} is devoted to the description of the Hamiltonian model used to represent the LiCN molecular system (Sec.~\ref{sec:system}), as well as to the description of the calculations to obtain the classical POs and a suitable Poincar\'e surface of section (Sec.~\ref{sec:trajec.calculations}), the quantum eigenenergies and eigenstates of the corresponding Hamiltonian operator (Sec.~\ref{sec:eigen.calculations}), and also the Lagrangian descriptors used to obtain the Maslov index (Sec.~\ref{sec:LDs.calculations}). Section~\ref{sec:results.discussion} is devoted to the joint presentation and discussion of the obtained results. In Sec.~\ref{sec:quantum.results} the obtained values of the parameters determining the frontier of scars are listed, where certain linear correlation is found. Also, as a representative example, one of the cases in the frontier of scars is illustrated and discussed through the depiction of the scarred wavefunctions and the three scarring POs involved. In Sec.~\ref{sec:semiclassical.results} the semiclassical quantization of the three scarring POs is carried out. First, the Maslov index of each PO must be obtained (Sec.~\ref{sec:Maslov}), which is trivially obtained by counting the number of turning points in two of the three POs, while this non-rigorous method fails in the third PO. Hence, a novel straightforward method based on Lagrangian descriptors is introduced and used for the third PO. Then, the quantization is performed (Sec.~\ref{sec:quantization}), such that, taking advantage of the linear correlation found in Sec.~\ref{sec:quantum.results}, a continuous semiclassical frontier between order and chaos is obtained. Last, the paper is summarized and the conclusions reached are presented in Sec.~\ref{sec:conclusions}.

\section{\label{sec:system.calculations}System description and calculations}

\subsection{\label{sec:system}Hamiltonian model}

The system studied in this work corresponds to the vibrational dynamics of the most abundant isotopic combination of the lithium isocyanide molecule $^{7}$Li$^{12}$C$^{14}$N. Regarding the Hamiltonian model used, some remarks are in order. On the one hand, the rotational motion is not considered, i.e., the model will account for the purely vibrational motion of the molecule. On the other hand, since the C-N bond is much stronger than the interactions with the Li atom, an adiabatic decoupling of the corresponding degree of freedom is feasible, such that the C-N bond length can be fixed at its equilibrium value, i.e., the model will describe the relative motion of the Li atom and the CN group. These simplifications lead to a chemically realistic model of the LiCN molecule with only two degrees of freedom, which is suitable for our purposes.

Considering the above simplifications, the Li-CN molecular system will be modeled by means of the Hamiltonian function
\begin{equation}
\label{eq:Hamiltonian}
H = \frac{P_R^2}{2\mu_1} + \frac{P_\theta^2}{2}\left( \frac{1}{\mu_1R^2} + \frac{1}{\mu_2r_\text{eq}^2} \right) + V(R,\theta),
\end{equation}
where $\mu_1 = m_\text{Li}(m_\text{C}+m_\text{N})/(m_\text{Li}+m_\text{C}+m_\text{N})$ and $\mu_2 = m_\text{C}m_\text{N}/(m_\text{C}+m_\text{N})$ are reduced masses ($m_\text{Li}$, $m_\text{C}$, and $m_\text{N}$ being the corresponding atomic masses), $r_\text{eq}=2.19$ a.u. is the fixed N-C equilibrium length, $R$ is the length between the CN group center of mass and the Li atom, and $\theta$ is the angle formed by the corresponding $r_\text{eq}$ and $R$ directions (i.e., N$\rightarrow$C and $\genfrac{}{}{0pt}{}{\text{C}}{\text{N}}$$\rightarrow$Li, respectively). Thus, e.g., $\theta=0$ corresponds to the linear configuration Li-CN, and $\theta=\pi$ rad to the linear configuration CN-Li. Last, $P_R$ and $P_\theta$ are the conjugate momenta corresponding to $R$ and $\theta$ coordinates, respectively, and $V(R,\theta)$ is the potential energy function describing the interatomic interaction.

The potential energy function $V(R,\theta)$ is taken from the literature~\cite{LiCN.PES}. It presents two minima: a relative minimum at $(R,\theta)=(4.79,0)$ (a.u., $\pi$ rad) with $V=2281$ cm$^{-1}$, corresponding to the Li-CN isomer, and an absolute minimum at $(R,\theta)=(4.35,1)$ (a.u., $\pi$ rad) with $V=0$, corresponding to the most stable CN-Li isomer. Both minima are separated by a saddle at $(R,\theta)=(4.22,0.29)$ (a.u., $\pi$ rad) with $V=3455$ cm$^{-1}$. These three characteristic points can be connected by the minimum energy path (MEP), i.e., the path connecting all characteristic points along which the variation of energy is minimal. Notice that, accordingly to the physics of the Li-CN molecular system, the potential energy function $V(R,\theta)$ is periodic in the angular coordinate $\theta$, with period $2\pi$ rad, and has a symmetry line at each value $\theta=k\pi$ rad $(k=0,\pm1,\pm2,\ldots)$. Finally, it is worth noting that the well around the absolute minimum (CN-Li isomer) is very anharmonic and, consequently, the transition from regular classical motion to chaos in this system~\cite{Arranz.LiCN.probability.chaos,Arranz.LiCN.bifurcation.chaos} takes place for energies around 1700 cm$^{-1}$, well below the isomerization barrier energy of 3455 cm$^{-1}$.

\subsection{\label{sec:trajec.calculations}Classical trajectories}

Classical trajectories will be calculated by numerically integrating the canonical equations of motion corresponding to the Hamiltonian function in Eq.~(\ref{eq:Hamiltonian}), where standard numerical methods will be used for the integration.

The scarring POs involved in the frontier of scars will be obtained by means of the systematic method described in Ref.~\cite{Arranz.LiCN.bifurcation.chaos}, which is based on the propagation of the symmetry line at $\theta=\pi$ rad.

Moreover, in order to get an graphical representation of the trajectories in phase space, which shows the different regular and chaotic regions, a suitable Poincaré surface of section (PSS) will be defined. For this purpose, the following canonical transformation will be applied
\begin{gather}
\label{eq:PSS.coordinates}
\rho = R - R_\text{eq}(\theta), \quad P_\rho = P_R, \nonumber\\
\vartheta = \theta, \quad P_\vartheta = P_\theta + P_R \frac{dR_\text{eq}(\theta)}{d\theta},
\end{gather}
where $R_\text{eq}(\theta)$ is a series expansion in $\theta$ coordinate that fits the MEP. Thus, for a given energy $E$, the PSS along the MEP will be defined in $(\vartheta,P_\vartheta)$ coordinates by taking $\rho=0$ and choosing an arbitrary branch (the negative one in our case) in the second degree equation for $P_\rho$ that arises from the Hamiltonian conservation condition $H(\rho,\vartheta,P_\rho,P_\vartheta)=E$.

\subsection{\label{sec:eigen.calculations}Eigenenergies and eigenstates}

In order to calculate the eigenenergies and eigenstates of the Hamiltonian operator corresponding to the Hamiltonian function in Eq.~(\ref{eq:Hamiltonian}), the {\em discrete variable representation-distributed Gaussian basis} (DVR-DGB) method proposed by Ba\v{c}i\'{c} and Light~\cite{DVR-DGB} will be used. As shown by these authors, the DVR-DGB method provides good accuracy for highly excited vibrational states, performing very well for the Li-CN molecular system, which was used as test system in their paper.

It is relevant to note here that, as a consequence of the separation of variables procedure for obtaining the vibrational Hamiltonian operator from the total Hamiltonian operator (see Ref.~\cite{DVR-DGB} and references therein), the angular coordinate is defined in the range $\theta\in[0,\pi]$ rad, within which consecutive quantum numbers $n_2=0,1,2,\ldots$ can be assigned, where $n_2$ represents excitation in the $\theta$ coordinate. However, in order to implement a suitable correspondence with classical mechanics, the range of the angular coordinate will be extended to $\theta\in[0,2\pi]$ rad by applying the symmetry line $\theta=\pi$ rad, such that only even quantum numbers $n_2=0,2,4,\ldots$ can be assigned. Due to this approach, the symmetry line $\theta=\pi$ rad plays a singular role in the procedure of semiclassical quantization, as discussed in Sec.~\ref{sec:Maslov}. Notice that, however, no restrictions apply to the quantum numbers corresponding to the radial coordinate, since in this case consecutive quantum numbers $n_1=0,1,2,\ldots$ can be assigned, where $n_1$ represents excitation in the $R$ coordinate.

Moreover, it has been shown in the literature~\cite{Arranz.LiCN.hbar.correlation1,Arranz.LiCN.hbar.correlation2} that by expanding the range of $\hbar$ values in the calculations, thus obtaining a correlation diagram of eigenenergies versus Planck constant, a conspicuous series of quantum resonances formed by broad ACs is observed, which constitutes the frontier that separates the regions of order and chaos in the Li-CN molecular system. In this paper we will calculate the position od this quantum frontier, where scarring phenomena first appear, demonstrating that it can be obtained by semiclassical quantization of the corresponding scarring POs.

In this way, the DVR-DGB method will be used at values $\hbar=\{0.01,0.02,\ldots,3.00\}\ \text{a.u.}$, obtaining the 130 low lying eigenstates for each value of $\hbar$ with eigenenergies converged to within 1~cm$^{-1}$. It is worth noting that, in order to maintain accuracy, the number of {\em rays} (the fixed values of $\theta$-coordinate taken in DVR-DGB method) must be increased as $\hbar$ decreases. Thus, a final basis set of 414-418 ray eigenvectors lying in 45 rays will be used in the range $\hbar\in[1.01,3.00]\ \text{a.u.}$, a basis set of 820-841 ray eigenvectors lying in 90 rays in the range $\hbar\in[0.31,1.00]\ \text{a.u.}$, and a basis set of 1480-1710 ray eigenvectors lying in 180 rays in the range $\hbar\in[0.01,0.30]\ \text{a.u.}$

\subsection{\label{sec:LDs.calculations}Lagrangian descriptors}

Introducing a novel method, Lagrangian descriptors will be used to compute the POs Maslov index, which are necessary for their semiclassical quantization. Lagrangian descriptors have been shown to be a very powerful tool to unveil the intricate invariant structures of the phase space of chaotic dynamical systems. Note that different definitions for the Lagrangian descriptors can be used, each of them leading to slightly different results~\cite{Mancho.LDs.different_definitions,Mancho.LDs.theoretical_proofs}. In our case, we will use the definition that has been shown to be suitable for the Li-CN molecular system, in particular in Refs~\cite{Revuelta.LDs.LiCN.1,Revuelta.LDs.LiCN.2}.

For a system with $N$ dimensions, the Lagrangian descriptors $M$ are defined as follows,
\begin{equation}
\label{eq:LDs.definition}
M_\pm({\bf z}_0;\alpha,\tau) = \pm\sum_{k=1}^{2N}\int_0^{\pm\tau}|\dot{z}_k(t)|^\alpha\,\text{d}t,
\end{equation}
where ${\bf z}=(z_1,\ldots,z_{2N})$ is the vector formed by the $N$ position variables and their corresponding $N$ conjugate momenta, such that, Lagrangian descriptors are a function depending on the initial condition ${\bf z}_0=(z_{1_0},\ldots,z_{2N_0})$, at time $t=0$, and two fixed parameters, the exponent $\alpha\in(0,1]$ and the integration time $\tau\in(0,+\infty)$. Note that, in the case of an unstable PO, backward $M_{-}$ and forward $M_{+}$ forms in Eq.~(\ref{eq:LDs.definition}) will permit to obtain the unstable and stable invariant manifolds, respectively. The overall Lagrangian descriptors $M$, as commonly used in the literature, are given by the sum of both forms, namely, $M=M_{-}+M_{+}$.

For the Li-CN molecular system, we have $N=2$ and ${\bf z}=(R,\theta,P_R,P_\theta)$. Additionally, we will take the value $\alpha=1$ for the exponent, which corresponds to the integration of the so-called {\em taxicab} norm~\cite{Krauser.taxicab.geometry} of the Hamiltonian flow $\dot{\bf z}(t)$ in Eq.~(\ref{eq:LDs.definition}), and the value $\tau=486$ fs for the integration time, which is large enough compared with the inverse of the stability exponent of the PO under study (namely, $|\lambda^{-1}|=87.50$ fs) as prescribed in Ref.~\cite{Revuelta.LDs.LiCN.2}. In any case, note that the choice of these values is heuristic, and then it is necessary to probe with different guesses until obtaining the clearest picture of the invariant manifolds.

In order to calculate the Maslov index of a PO, different initial conditions ${\bf z}_0=(R_0,\theta_0,P_{R_0},P_{\theta_0})$ will be taken along the PO in configuration space, exploring the energetically accessible momentum space at each position, as described in the discussion of the results in Sec.~\ref{sec:Maslov}.

\section{\label{sec:results.discussion}Results and discussion}

\subsection{\label{sec:quantum.results}Quantum results}

Although an extensive correlation diagram within the ranges given in Sec.~\ref{sec:eigen.calculations} for $\hbar$ and the corresponding eigenenergies has been calculated, we will mainly focus on the region where the quantum transition from order to chaos occurs, i.e., the frontier of scars. The whole correlation diagram is not shown here, but it is reported in the previous article~\cite{Arranz.LiCN.correspondence.resonances}. Instead, a magnification centered on the frontier of scars, with the semiclassical results superimposed, will be shown below in Fig.~\ref{fig:correlation.diagram}.

\begin{table}[b]
\caption{\label{tab:scars.frontier.data}%
Values of the parameters determining the series of avoided crossings that constitutes the frontier of scars separating the regions of order and chaos in the correlation diagram of eigenenergies versus Planck constant. For each avoided crossing, the quantum number $n$, the Planck constant value $\hbar_n$, the lower $E_n^-$ and upper $E_n^+$ eigenenergy values, and their corresponding state numbers, $N_n^-$ and $N_n^+$, are listed.}
\begin{ruledtabular}
\begin{tabular}{cccccc}
$n$ & $\hbar_n$ (a.u.) & $E_n^-$ (cm$^{-1}$) & $N_n^-$ & $E_n^+$ (cm$^{-1}$) & $N_n^+$ \\
\hline
12 & 2.430 & 3601 & 10 & 3694 & 11 \\
14 & 1.930 & 3205 & 11 & 3274 & 13 \\
16 & 1.600 & 2944 & 13 & 3000 & 14 \\
18 & 1.370 & 2766 & 16 & 2814 & 17 \\
20 & 1.200 & 2638 & 18 & 2680 & 19 \\
22 & 1.062 & 2526 & 21 & 2564 & 22 \\
24 & 0.955 & 2443 & 24 & 2479 & 25 \\
26 & 0.867 & 2374 & 27 & 2407 & 29 \\
28 & 0.794 & 2318 & 31 & 2348 & 32 \\
30 & 0.733 & 2272 & 35 & 2301 & 36 \\
32 & 0.684 & 2242 & 39 & 2270 & 40 \\
\end{tabular}
\end{ruledtabular}
\end{table}
The values of the parameters determining the position of the series of ACs that constitutes the frontier of scars are listed in Table~\ref{tab:scars.frontier.data}. The center point of each AC, given by the corresponding value of the Planck constant $\hbar_n$, is defined as the value of $\hbar$ at which the coupling $\langle\psi_i|\partial/\partial\hbar|\psi_j\rangle$ between the two eigenstates $|\psi_i\rangle$ and $|\psi_j\rangle$ involved in the AC reaches its maximum. It is worth noting that the mixing between both states is completely determined by the coupling $\langle\psi_i|\partial/\partial\hbar|\psi_j\rangle$~\cite{Arranz.LiCN.hbar.correlation1,Arranz.LiCN.hbar.correlation2}. Accordingly, the lower $E_n^-$ and upper $E_n^+$ eigenenergy values in Table~\ref{tab:scars.frontier.data} correspond to the energy of the two eigenstates involved in the AC at Planck constant $\hbar_n$. Also, the corresponding state numbers $N_n^-$ and $N_n^+$ are listed, where $N=1$ stands for the ground state. Observe that, due to the existence of ACs where the interaction of the involved states is ostensibly small, this giving rise to very sharp ACs non-observable by naked-eye inspection in the eigenenergies correlation diagram, the state numbers $N_n^-$ and $N_n^+$ are not consecutive in all cases. Notice that the well known non-crossing rule ensures that, in this system, all eigenstates undergo ACs.

All parameters in Table~\ref{tab:scars.frontier.data} are labeled by the quantum number $n$ associated with each AC, which is obtained from the nodal pattern of the corresponding scarred wavefunction. Namely, the quantum number $n$ is calculated by counting the number of times that the graph of the scarring PO crosses a nodal line of the scarred wavefunction. Note that, due to the extended range of the angular coordinate $\theta\in[0,2\pi]$ rad mentioned above in Sec.~\ref{sec:eigen.calculations}, all values of $n$ are even numbers. The quantum number $n$ represents excitation neither in the $R$ coordinate nor in the $\theta$ coordinate, but in the coordinate defined along the corresponding scarring PO. As was shown in Ref.~\cite{Polavieja.scars.in.groups}, this quantum number of scarred states is related to the different bands appearing in the appropriate low-resolution spectrum, such that each band in the spectrum is associated with the corresponding scarred state and its quantum number. Moreover, the low-resolution spectrum of the Li-CN molecular system, related to the scarred states involved in the frontier of scars, was studied in Ref.~\cite{Arranz.scars.low-res.spectrum}.

\begin{figure}[t]
\includegraphics{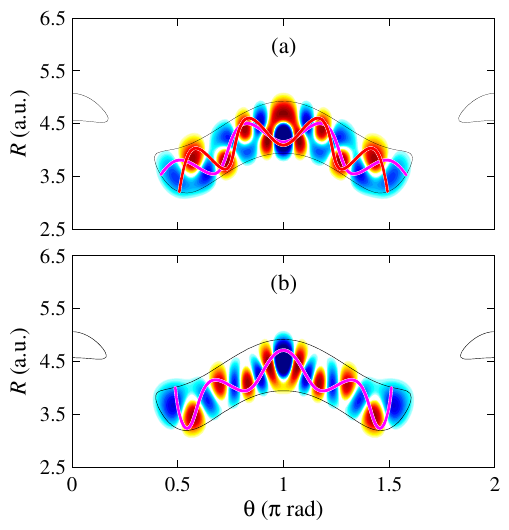}
\caption{\label{fig:wavefunctions}%
Scarred wavefunctions corresponding to the upper~(a) and lower~(b) states involved in the avoided crossing with quantum number $n=16$. The scarring periodic orbits, depicted in thick line, have been superimposed on the wavefunctions. The energy contour corresponding to each eigenenergy has also been included, depicted in thin line.}
\end{figure}
\begin{figure}[t]
\includegraphics{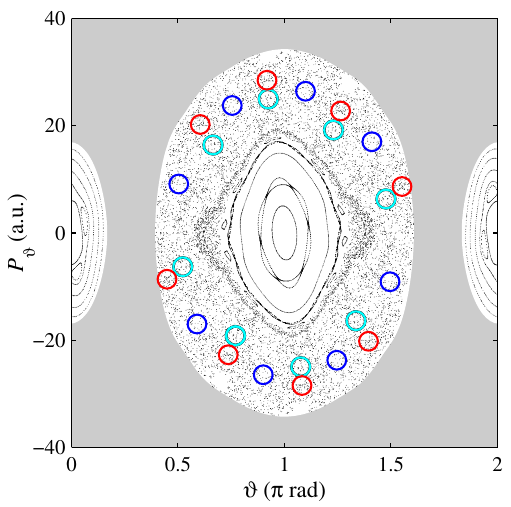}
\caption{\label{fig:PSS}%
Composite Poincar\'e surface of section, defined along the minimum energy path, for energy  $E=2972\text{ cm}^{-1}$, which corresponds to the middle energy at the avoided crossing with quantum number $n=16$. The periodic orbits referred to in the text as PO-A, PO-B, and PO-C are marked with red (dark), cyan (light), and blue (darkest) open circles ($\bigcirc$), respectively. Gray region represents the energetically forbidden region.}
\end{figure}
As a representative example, the case of the AC corresponding to the quantum number $n=16$ is shown in Fig.~\ref{fig:wavefunctions}, where the scarred wavefunctions are depicted with their scarring POs superimposed on them. Observe that, characterizing the scarring phenomenon, the extremes (maxima and minima) of the wavefunctions are distributed along the corresponding POs, such that counting the number of times that the graph of a PO crosses a nodal line of the scarred wavefunction, the quantum number $n=16$ is obtained. Strictly speaking, the scarring POs must be {\em isolated unstable} POs, otherwise we would have a localization phenomenon rather than a scarring phenomenon. In order to show the isolated and unstable character of the involved POs, we have depicted in Fig.~\ref{fig:PSS} a composite PSS for the middle energy at the AC with quantum number $n=16$, where the periodic points corresponding to the involved POs have also been superimposed as open circles. As can be observed, in all cases each periodic point (at the center of its open circle mark) is immersed in the chaotic region, evidencing the isolated and unstable character of the corresponding PO. Additionally, the stability of the involved POs throughout the series of ACs have been determined by the calculation of the trace of the monodromy matrix.

For the $n=16$ case shown in Fig.~\ref{fig:PSS}, all of the three POs are unstable and isolated, however, this is not the case for all instances in the series of the ACs. For the lower state, only one PO is involved (hereafter referred to as PO-C), which is unstable and isolated throughout the series. For the upper state, two POs are involved, one more and one less extended in the $\theta$-coordinate [approximately, $\theta\in[0.4,1.6]\ \pi$ rad and $\theta\in[0.5,1.5]\ \pi$ rad, respectively, in the case shown in Fig.~\ref{fig:wavefunctions}~(a)]. The less extended one (hereafter referred to as PO-B) is unstable and isolated throughout the series, while the more extended one (hereafter referred to as PO-A) is unstable and isolated for the ACs where $n=12,14,16,18,20$, but it is stable for the ACs where $n=22,24,26,28,30,32$. These facts will be taken into account again in Sec.~\ref{sec:quantization}, where the semiclassical quantization is discussed. In any case, throughout the series of ACs, there exists at least one isolated unstable PO scarring the corresponding eigenstate. Additional details about the characteristics of the POs in connection with the onset of chaos of the LiCN molecular system can be obtained in Ref.~\cite{Arranz.LiCN.bifurcation.chaos}.

\begin{figure}[t]
\includegraphics{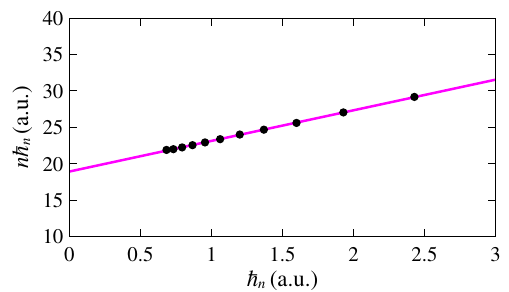}
\caption{\label{fig:hbar.frontier}%
Linear correlation in the frontier of scars. Shown is the quantization $n\hbar_n$ versus $\hbar_n$, $n$ and $\hbar_n$ being the quantum numbers and the Planck constant values, respectively, given in Table~\ref{tab:scars.frontier.data}. The least-squares fitting of a straight line to the data points is also depicted.}
\end{figure}

Moreover, it is interesting to note that the values of the Planck constant $\hbar_n$, i.e.,~those where the ACs are centered, have a high linear correlation with the quantized $n\hbar_n$, as quantitatively indicated by the Pearson correlation coefficient $r=0.99986$. This linear correlation is shown graphically in Fig.~\ref{fig:hbar.frontier}, where the least-squares fitting of a straight line to the data points is also depicted. The fitted straight line
\begin{equation}
\label{eq:n-hbarn.fitting}
n\hbar_n = a + b\,\hbar_n
\end{equation}
has intercept $a=18.92\pm0.03$ a.u. and slope $b=4.20\pm0.02$, with a mean squared error of $0.00166\text{ (a.u.)}^2$. This result will be used below in Sec.~\ref{sec:quantization} in order to define a continuous curve, derived from semiclassical quantization, determining the frontier between order and chaos.

\subsection{\label{sec:semiclassical.results}Semiclassical results}

The semiclassical quantization of an unstable PO, as obtained by Gutzwiller in the derivation of his trace formula~\cite{Gutzwiller.trace.formula}, is given by
\begin{equation}
\label{eq:S.quantization}
S = \hbar\left(n + \frac{\mu}{4}\right),
\end{equation}
where $S$ is the classical action over one period of the orbit,%
\footnote{Notice that the factor $(1/2\pi)$ have been included in the definition of the action, otherwise the original Planck constant $h$ should be used instead of the reduced Planck constant $\hbar$.}
$\hbar$ is the Planck constant, $n$ is the quantum number, and $\mu$ is the Maslov index of the PO. This index is an invariant of the PO, which counts the number of conjugate points over one period of the orbit. Then, in order to accomplish the quantization of the POs involved in the frontier of scars, it is previously required the calculation of their Maslov indices.

\subsubsection{\label{sec:Maslov}Maslov index}

As pointed out in Sec.~\ref{sec:eigen.calculations}, the original range of the angular coordinate in the Li-CN molecular system is $\theta\in[0,\pi]$ rad, such that, due to the symmetry of the system, the line $\theta=\pi$ rad behaves as a hard-wall potential, i.e., at $\theta=\pi$ rad the incident angle of a classical trajectory is equal to the reflected angle. The prescription of the semiclassical quantization when there is a hard-wall is to add a value of 2 to the Maslov index, accounting for the phase loss in the semiclassical propagation of the wave along the classical trajectory. When the range of the angular coordinate is extended to $\theta\in[0,2\pi]$ rad by applying the symmetry line $\theta=\pi$ rad, the prescription of adding a value of 2 to the Maslov index remains. Observe that this prescription is consistent with the fact that, in the extended range $\theta\in[0,2\pi]$ rad, all eigenfunctions are symmetric (with respect to the line $\theta=\pi$ rad), otherwise antisymmetric eigenfunctions, with odd quantum numbers, should also exist. Accordingly, a value of 2 must be added to the Maslov indices of the POs calculated in the extended range $\theta\in[0,2\pi]$ rad.

A rigorous calculation of the Maslov index can be implemented by means of different techniques, from the long-established method of Eckhardt and Wintgen~\cite{Maslov.winding.number}, based on the winding number of the invariant manifolds of the PO, to the most recent method of Vergel {\em et al.}~\cite{Maslov.geometrodynamic}, based on the number of zeros of the Jacobi field of the geodesic line corrersponding to the PO in the geometrodynamic approach. In all cases a rigorous calculation requires a demanding mathematical work.

Moreover, in some cases the Maslov index of a PO can be obtained by means of an easy method, as is the counting of the number of turning points in each degree of freedom over one period of the orbit. Thus, for example, if we take Ref.~\cite{Maslov.geometrodynamic}, where Maslov indices are calculated in the framework of the geometrodynamic approach, and we focus on the unstable POs of the two-dimensional system represented in Fig.~9 of this reference, the counting of the number of turning points in the four cases represented in panels (b)-(e), which correspond to POs with more and less complex graphs, gives the correct Maslov indices (listed in Table~II of Ref.~\cite{Maslov.geometrodynamic}). However, the counting of the number of turning points in the two cases represented in panels (a) and (f), which correspond to POs with extremely simple graphs (oblique and horizontal straight lines, respectively), gives wrong Maslov indices. The case in panel (a), the oblique straight line graph, is rather trivial. The correct Maslov index is 2 and the counting of the number of turning points is 4 (2 in each degree of freedom), but a coordinate rotation leading from oblique to either horizontal or vertical straight line yields a counting of 2 turning points. On the contrary, the case in panel (f), the horizontal straight line graph, is amazing. Indeed, as in the previous case, we would expect a Maslov index of 2, however the correct Maslov index is 16. This case exemplify the complex behavior that the Maslov index can sometimes exhibit, such that easy methods as the counting of the number of turning points should only be used when the obtained results can be tested.

\begin{figure}[t]
\includegraphics{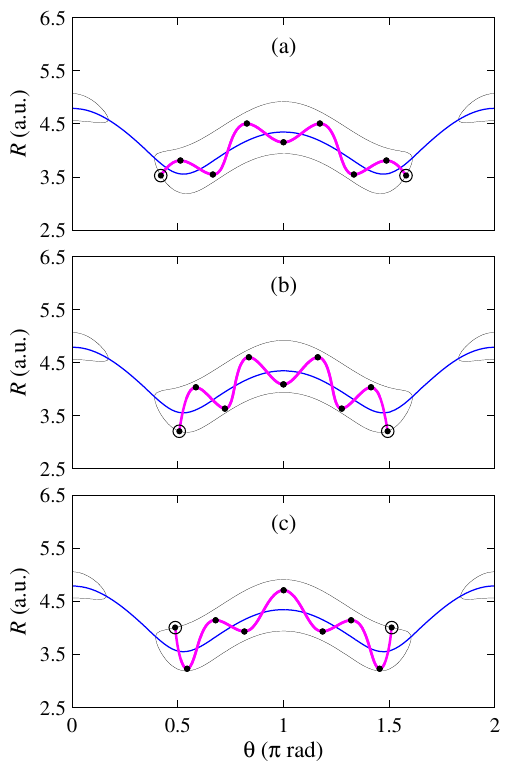}
\caption{\label{fig:PO.turning.points}%
Turning points of the scarring periodic orbits associated with the scarred wavefunctions corresponding to the upper~[(a) and (b)] and lower~(c) states involved in the series of ACs that constitutes the frontier of scars. These periodic orbits are referred to in the text as PO-A (a), PO-B (b), and PO-C (c). The turning points in the radial coordinate $R$ are marked with dots ($\bullet$) while the turning points in the angular coordinate $\theta$ are marked with open circles ($\bigcirc$). The graph of each periodic orbit, the minimum energy path, and the corresponding energy contour are represented by thick magenta, medium blue, and thin black lines, respectively.}
\end{figure}
Returning to the Li-CN molecular system, the graph of the three scarring POs involved in the frontier of scars have been depicted separately in Fig.~\ref{fig:PO.turning.points}, where their turning points in each degree of freedom are highlighted. Observe that by following each path over one period, i.e., going and coming back to the initial point, the number of turning points is the same in the three cases. The number of turning points in the radial coordinate $R$, i.e., the points where its conjugate momentum takes the value $P_R=0$ changing the sign, is 16. Also, the number of turning points in the angular coordinate $\theta$, i.e., the points where its conjugate momentum takes the value $P_\theta=0$ changing the sign, is 2. Then, accounting the value 2 due to the hard-wall line at $\theta=\pi$ rad, the Maslov index obtained by counting the number of turning points in each degree of freedom is $\mu=16+2+2=20$. As we will see in Sec.~\ref{sec:quantization}, the semiclassical energies obtained from the quantization with the Maslov index $\mu=20$ are consistent with the eigenenergies obtained from the quantum calculations for both PO-A and PO-B, i.e., those associated with the scarred wavefunctions corresponding to the upper states in the series of ACs. However, the results are inconsistent for PO-C, i.e., that one associated with the scarred wavefunctions of the lower states in the series of ACs, suggesting that the Maslov index $\mu=20$ is not correct in this case.

In order to obtain the Maslov index for PO-C, we will introduce a novel method, based on the rigorous (and mathematically demanding) technique of Eckhardt and Wintgen~\cite{Maslov.winding.number}, but making it easy through the use of Lagrangian descriptors. Eckhardt and Wintgen showed that the Maslov index of an unstable PO is given by the number of half-turns around the PO of the associated invariant manifolds over one period. When this calculation is implemented in configuration space, rather than in phase space, the  existence of simultaneous turning points (i.e., points of the trajectory where all momentum values vanish at the same time value) must be taken into account. This is the case when the path of the PO in configuration space is self-retracing. Due to the singularities that appear at the simultaneous turning points in configuration space, the calculation fails at these points. The solution, however, is straightforward: The value 1 must be added to the number of half-turns for each simultaneous turning point. In any case, the calculation of the number of half-turns is mathematically demanding. The novelty in our method is to calculate the number of half-turns of the invariant manifolds by means of a suitable graphical representation of the (easy to calculate) Lagrangian descriptors along the PO.

\begin{figure*}[t]
\includegraphics{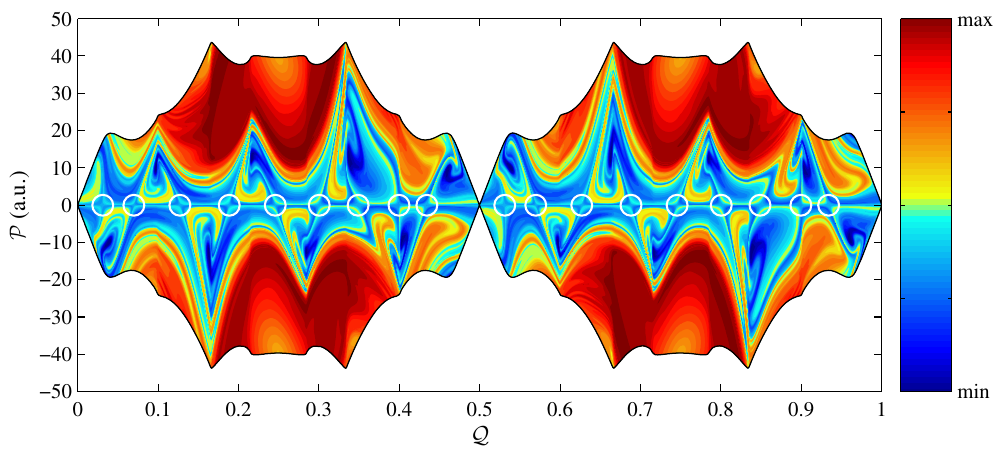}
\caption{\label{fig:LD_maslov}%
Color scale representation of the forward form $M_+({\cal Q,P})$ of the Lagrangian descriptors calculated along the unstable periodic orbit in Fig.~\ref{fig:PO.turning.points}~(c), referred to in the text as PO-C, over one period. Colorless white area represents energetically inaccessible region. The crossings of the stable invariant manifold with the line ${\cal P}=0$ are marked with white open circles ($\bigcirc$).}
\end{figure*}
Lagrangian descriptors have been shown to be a straightforward tool to depict the invariant manifolds of isolated POs embedded in the chaotic region of nonlinear systems, in particular also in the Li-CN molecular system~\cite{Revuelta.LDs.LiCN.1,Revuelta.LDs.LiCN.2}. In these works, the Lagrangian descriptors were calculated in a typical PSS representation. In our case, and for the sake of computing Maslov indices from them, a little different surface of section will be used. Thus, for a given total energy, we will consider a surface of section along the PO in configuration space, parameterizing the position coordinates $(R,\theta)$ by means of the normalized length of the path, ${\cal Q}$, such that ${\cal Q}=0$ corresponds to the left simultaneous turning point, ${\cal Q}=0.5$ corresponds to the right simultaneous turning point, and ${\cal Q}=1$ again corresponds to the left simultaneous turning point. Moreover, at each position ${\cal Q}$ in configuration space, all energetically accessible momentum values will be explored, parameterizing the momentum coordinates $(P_R,P_\theta)$ by means of the form
\begin{equation}
\label{eq:P.LDs}
{\cal P} = \left( \phi - \phi^\text{PO} \right) \| {\bf P} \|_\phi,
\end{equation}
where $\| {\bf P} \|_\phi$ and $\phi\in\left[\phi^\text{PO}-\pi,\phi^\text{PO}+\pi\right]$ rad are the modulus and angle, respectively, of the vector ${\bf P}=(P_R,P_\theta)$ in momentum space, $\phi^\text{PO}$ being the angle of the momentum corresponding to the PO for each given position ${\cal Q}$. Observe that, as follows from the form of the Hamiltonian function in Eq.~(\ref{eq:Hamiltonian}), the curve of the energetically accessible values in momentum space (for a given position ${\cal Q}$) is a ellipse rather than a circle. Consequently, the modulus $\| {\bf P} \|_\phi$ depends on the angle $\phi$, hence the notation used. In this way, for a given PO, the initial condition ${\bf z}_0=(R_0,\theta_0,P_{R_0},P_{\theta_0})$ of the Lagrangian descriptors $M_\pm({\bf z}_0)$ will be given by the parameterized position coordinates $(R_0,\theta_0)=(R,\theta)_{\cal Q}$ and the parameterized momentum coordinates $(P_{R_0},P_{\theta_0})=(P_R,P_\theta)_{\cal Q,P}$, such that the initial condition ${\bf z}_0$, as well as the Lagrangian descriptors $M_\pm({\bf z}_0)$, will be a function of $({\cal Q,P})$. Finally, in order to avoid a double counting of the number of half-turns around the PO of the associated invariant manifolds, only one of the two (either the stable or the unstable) invariant manifolds will be obtained by taken either forward $M_+({\cal Q,P})$ or backward $M_-({\cal Q,P})$ form.

The forward form $M_+({\cal Q,P})$ calculated for PO-C is depicted in Fig.~\ref{fig:LD_maslov}. On the one hand, throughout the range of ${\cal Q}$ we can observe a horizontal line for ${\cal P}=0$, which corresponds exactly to the PO path. In fact, this line is the locus were stable and unstable invariant surfaces intersect at the origin of the tangent space of the PO. Consequently, this horizontal line should evidently appear in the Lagrangian descriptors. On the other hand, we can observe a series of apparently different lines crossing the line ${\cal P}=0$, which are approximately straight in the neighborhood of the crossings while they stretch and twist as recede from these ones. These crossing lines correspond to the two branches of the stable invariant manifold, hence they are not different lines but a single line which is given by the intersection of the stable invariant manifold and our surface of section defined above. The aforementioned stretching and twisting, resulting from the nonlinear character of the Li-CN molecular system, hinder the visualization of this geometric object as a single line. Note that, since the value ${\cal P}=0$ corresponds to the momentum of the PO, at each crossing point the corresponding branch of the invariant manifold coincides with the PO, i.e., it determines a turn of the branch around the PO. Therefore, by counting the successive crossing points we are counting the alternating turns of each branch, namely, we are counting the number of half-turns of the invariant manifold around the PO.

In this way, the correct Maslov index for PO-C would be obtained as follows. The number of half-turns of the invariant manifold around the PO over one period, calculated by counting the number of crossings of the invariant manifold with the line ${\cal P}=0$ in Fig.~\ref{fig:LD_maslov}, is 18. The number of simultaneous turning points of the PO in the configuration space, which are clearly represented in Fig.~\ref{fig:LD_maslov} as the two singularities at ${\cal Q}=0,1$ and ${\cal Q}=0.5$, is 2. Last, the hard-wall line at $\theta=\pi$ rad adds the value 2. Eventually, the Maslov index of the third PO will be $\mu=18+2+2=22$.

\subsubsection{\label{sec:quantization}Quantization}

\begin{figure}[t]
\includegraphics{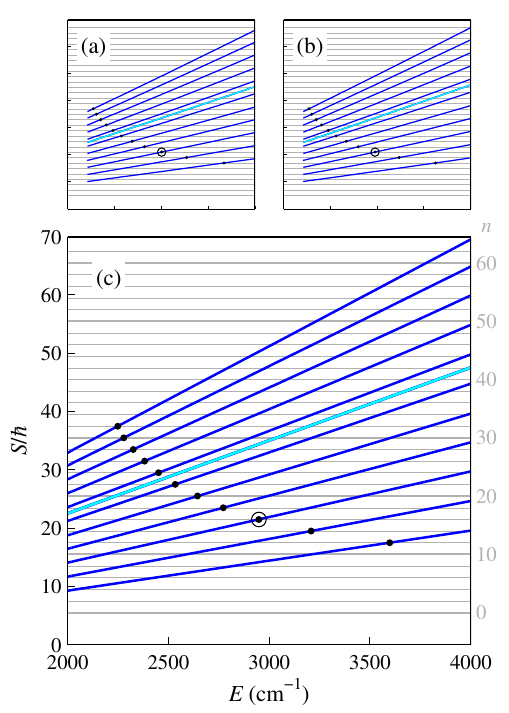}
\caption{\label{fig:PO.quantization}%
Quantization of the scarring periodic orbits associated with the scarred wavefunctions corresponding to the upper~[(a) and (b)] and lower~(c) states involved in the series of avoided crossings that constitutes the frontier of scars. These periodic orbits are referred to in the text as PO-A (a), PO-B (b), and PO-C (c). The classical action $S(E)$ obtained from Eq.~(\ref{eq:S.classical}), in the dimensionless form $S(E)/\hbar$ (with $\hbar=1$ a.u.), is depicted in cyan (light) line. The dimensionless classical action for each avoided crossing $S(E)/\hbar_n$ (with $\hbar_n$ values given in Table~\ref{tab:scars.frontier.data}), is depicted in blue (dark) line. Horizontal gray lines represent the quantization condition in Eq.~(\ref{eq:S.quantization}) for the indicated quantum number $n$. The points where the quantization condition is satisfied for each avoided crossing are marked with dots ($\bullet$). The open circle ($\bigcirc$) marks the representative case with quantum number $n=16$ discussed in Sec.~\ref{sec:quantum.results}. The axis in little panels (a) and (b) are the same as in the big panel (c).}
\end{figure}

In order to quantize the three scarring POs involved in the frontier of scars, the quantization condition in Eq.~(\ref{eq:S.quantization}) has been applied, taking in each case the corresponding Maslov index calculated above, and calculating the classical action $S$ as follows,
\begin{align}
\label{eq:S.classical}
S &= \frac{1}{2\pi}\oint_\text{PO}{\bf P}\cdot\text{d}{\bf Q}
\nonumber\\
  &= \frac{1}{2\pi}\int_0^T\left(P_R\dot{R} + P_\theta\dot{\theta}\right)\text{d}t
\nonumber\\
  &= \frac{1}{2\pi}\left(\frac{1}{\mu_1}\int_0^T\!\!\!\!P_R^2\,\text{d}t
      + \frac{1}{\mu_1}\int_0^T\!\!\frac{P_\theta^2}{R^2}\,\text{d}t
      + \frac{1}{\mu_2r_\text{eq}^2}\int_0^T\!\!\!\!P_\theta^2\,\text{d}t\right),
\end{align}
where the Hamiltonian function of the system [Eq.~(\ref{eq:Hamiltonian})] and the canonical equations of Hamilton have been used. Note that the integration is performed over the period $T$ of the corresponding PO. Also note that the obtained action depends on the energy of the PO, such that, by integrating in Eq.~(\ref{eq:S.classical}) for different energy values, the graph of the function $S(E)$ is obtained. This graph, in the dimensionless form $S(E)/\hbar$ (with $\hbar=1$ a.u.), is depicted in Fig.~\ref{fig:PO.quantization} for the three POs, namely PO-A, PO-B, and PO-C. By taking the values of the Planck constant $\hbar_n$ corresponding to each AC with quantum number $n$ (values given in Table~\ref{tab:scars.frontier.data}), the dimensionless action $S(E)/\hbar_n$ for each AC, also depicted in Fig.~\ref{fig:PO.quantization}, is obtained. Thus, the quantization condition in Eq.~(\ref{eq:S.quantization}) can be written in the form
\begin{equation}
\label{eq:Soverhbar.quantization}
\frac{S(E_n)}{\hbar_n} = \left(n + \frac{\mu}{4}\right),
\end{equation}
where $E_n$ is the quantized energy corresponding to the AC with quantum number $n$. In Fig.~\ref{fig:PO.quantization}, the left hand side in Eq.~(\ref{eq:Soverhbar.quantization}) is represented by the blue (dark) lines, whilst the right hand side corresponds to the horizontal gray lines, where the value $\mu=20$ has been taken in panels (a) and (b) [cases PO-A and PO-B, respectively], and the value $\mu=22$ in panel (c) [case PO-C].

\begin{figure*}[t]
\includegraphics{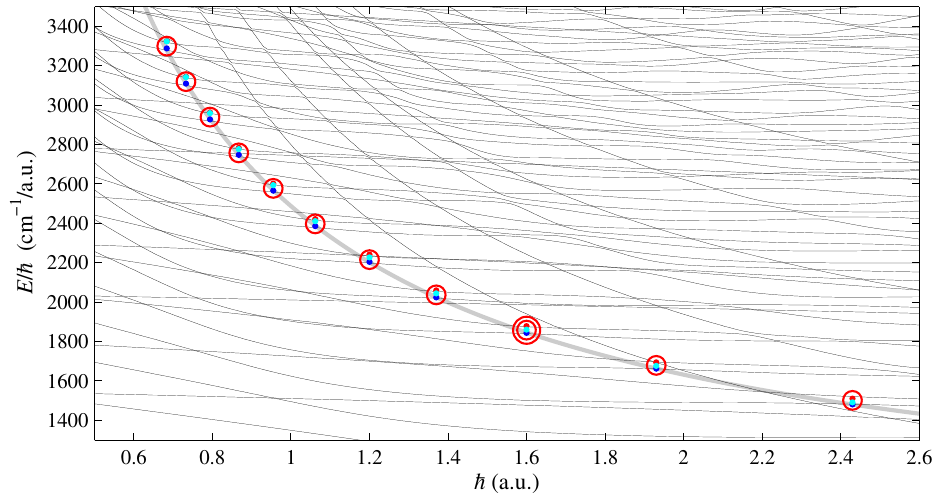}
\caption{\label{fig:correlation.diagram}%
Magnification of the correlation diagram of eigenenergies versus Planck constant centered on the quantum frontier of scars [marked with red open circles ($\bigcirc$)], that separates the regions of order and chaos. The circled circle symbol ($\circledbigcirc$) marks the representative case with quantum number $n=16$ discussed in Sec.~\ref{sec:quantum.results}. On grounds of graphical clarity, energy is divided by Planck constant. The energies obtained from the semiclassical quantization of the three periodic orbits in Fig.~\ref{fig:PO.turning.points}~(a) PO-A, (b) PO-B, and (c) PO-C are marked with red (dark), cyan (light), and blue (darkest) dots ($\bullet$), respectively. The semiclassical frontier between order and chaos is depicted in thick gray line.}
\end{figure*}
The quantized energies obtained from Eq.~(\ref{eq:Soverhbar.quantization}) are superimposed on the correlation diagram of eigenenergies versus Planck constant in Fig.~\ref{fig:correlation.diagram}, where the series of broad ACs that constitutes the frontier of scars has been marked with open circles. Notice that, from right to left in the figure, the quantum number $n$ increases from $n=12$ to $n=32$ across the series, as indicated in Table~\ref{tab:scars.frontier.data} where all parameters determining the series are listed. As was mentioned in Sec.~\ref{sec:quantum.results}, the figure shows how the state numbers $N_n^-$ and $N_n^+$ are not consecutive in cases $n=14$ and $n=26$, due to the existence of sharp ACs. Notice also that the frontier of scars actually separates the region of order (below the frontier), characterized by sharp ACs, and the region of chaos (above the frontier), characterized by overlapping ACs that lead to level repulsion property. However, since the Li-CN molecular system exhibits mixed chaos, such that it can be observed the existence of stability islands embedded in the classical chaotic sea, then it can also be observed the existence of sharp ACs embedded in the quantum level repulsion sea.

Moreover, observe that the semiclassical results are in good agreement with quantum results, demonstrating that the frontier of scars can be obtained by means of the semiclassical quantization of the corresponding scarring POs. More specifically, it can be observed a good agreement throughout the series of ACs for cases PO-A and PO-C, while a rough agreement is observed at the beginning of the series ($n=12$), which progressively becomes a good agreement as $n$ increases, for the case PO-B. It is interesting to note that, as was mentioned in Sec.~\ref{sec:quantum.results}, PO-A is unstable for quantum numbers $n=12,14,16,18,20$, which are the quantum numbers for which the agreement for PO-B is not good enough, whilst it is stable for quantum numbers $n=22,24,26,28,30,32$, which are the quantum numbers for which the agreement is good enough. In other words, it seems that, when quantum number increases from $n=20$ to $n=22$, PO-B replaces PO-A in the role of {\em isolated unstable} PO required for scarring phenomena.

Finally, we will take advantage of the linear relationship in Eq.~(\ref{eq:n-hbarn.fitting}) to obtain a semiclassical continuous expression for the frontier between order and chaos. Notice that from Eq.~(\ref{eq:n-hbarn.fitting}) we can write
\begin{equation}
\label{eq:hbarn.vs.n}
\hbar_n = \frac{S_\infty}{\left(n - n_0\right)},
\end{equation}
where $S_\infty=18.92\pm0.03$ a.u. and $n_0=4.20\pm0.02$ are the intercept $a$ and slope $b$, respectively, in Eq.~(\ref{eq:n-hbarn.fitting}). The physical meaning of parameter $n_0$ in Eq.~(\ref{eq:hbarn.vs.n}) is clear: Since $\lim_{n\to n_0}\hbar_n=+\infty$, it is the lower limit of the open interval defining the domain, i.e., $n\in(n_0,+\infty)\mid n\in2\mathbb{N}$. In other words, the quantum number $n$ could take any even value strictly greater than $n_0$. Moreover, the physical meaning of parameter $S_\infty$ is obtained by inserting Eq.~(\ref{eq:hbarn.vs.n}) into the quantization condition in Eq.~(\ref{eq:Soverhbar.quantization}), namely
\begin{align}
\label{eq:Sn.vs.n}
S_n &= \hbar_n\left(n + \frac{\mu}{4}\right) \nonumber\\
    &= S_\infty\frac{\left(n + \mu/4 \right)}{\left(n - n_0\right)},
\end{align}
where it is fulfilled that $\lim_{n\to\infty}S_n=S_\infty$. Then, the parameter $S_\infty$ is the asymptotic value of the action in the semiclassical limit $n\to\infty$ (i.e., $\hbar_n\to0$). Notice that Eq.~(\ref{eq:hbarn.vs.n}) accurately retrieves the values of the Planck constant $\hbar_n$ for the corresponding quantum number $n$ at each AC listed in Table~\ref{tab:scars.frontier.data}, such that by applying the quantization condition, which is represented in Fig.~\ref{fig:PO.quantization}, the quantized energies $E_n$ depicted in Fig.~\ref{fig:correlation.diagram} are also accurately obtained. Therefore, if in this process we consider a continuous rather than discrete domain for the ``quantum'' number $n\in(n_0,+\infty)$ when applying the ``quantization'' condition, then an also continuous rather than discrete set for the energies will be obtained. This continuous set of energies constitutes the semiclassical frontier between order and chaos, which is depicted superimposed on the correlation diagram in Fig.~\ref{fig:correlation.diagram} for the case PO-C.

It is worth noting that the linear correlation shown in Fig.~\ref{fig:hbar.frontier} and the corresponding fitted straight line given in Eq.~(\ref{eq:n-hbarn.fitting}) should be an approximation, i.e., the behavior of $n\hbar_n$ versus $\hbar_n$ is nearly but not strictly linear. In particular, as the position of the point with highest quantum number ($n=32$) in Fig.~\ref{fig:hbar.frontier} seems to indicate, the series could deviate from the linear behavior as $n$ increases (i.e., $\hbar_n$ decreases). Consequently, the relationships in Eqs.~(\ref{eq:hbarn.vs.n}) and~(\ref{eq:Sn.vs.n}), both derived from Eq.~(\ref{eq:n-hbarn.fitting}), will also be approximate. However, we think that their qualitative behavior, namely, monotonically decreasing functions (considering a continuous domain) with vertical asymptote at $n=n_0$ and horizontal asymptote at $\hbar=0$ or $S=S_\infty$ in each case, is the correct one. The value $n_0\approx4$ obtained from the fitting implies that the minimum value for the quantum number is $n=6$, since it must be an even integer strictly greater than $n_0$. However, the AC with the lowest possible quantum number, since there are not lower states leading to an AC, corresponds to $n=8$ (see the previous article~\cite{Arranz.LiCN.correspondence.resonances}), hence the vertical asymptote could be at $n=6$ (due to the deviation from the linear behavior) rather than at $n=4$. Moreover, the value $S_\infty\approx19$ a.u. obtained from the fitting corresponds to an energy around $E\approx1700$ cm$^{-1}$ in the classical action function $S(E)$ for PO-C. Note that, as was pointed out in Sec.~\ref{sec:system}, this energy value also corresponds to the classical transition from order to chaos in the Li-CN molecular system. However, if we assume the deviation from the linear behavior suggested by the point corresponding to $n=32$ in Fig.~\ref{fig:hbar.frontier}, then the horizontal asymptote should be at a value greater than the fitted parameter, such that there would be no direct relation to the threshold energy of transition to classical chaos. As a conjecture connected with the scarring phenomena, perhaps the horizontal asymptote could be at $S\approx22$ a.u., which corresponds to the energy $E=1958$ cm$^{-1}$ at which PO-C bifurcates becoming an isolated unstable PO. In any case, the question of the semiclassical limit $\hbar_n\to0$ of the series of ACs that constitutes the frontier of scars remains an open question.

\section{\label{sec:conclusions}Summary and conclusions}

We have studied the frontier of scars, previously established in the literature~\cite{Arranz.LiCN.hbar.correlation1,Arranz.LiCN.hbar.correlation2}, that separates the regions of order and chaos in the correlation diagram of eigenenergies versus Planck constant of the Li-CN molecular system, with the purpose of demonstrating that it can be obtained through the semiclassical quantization of the involved scarring POs. It should be remarked that, as shown by previous work of our group, this method is like a microscope in the phase space, where by changing the magnification power by decreasing the value of $\hbar$, many relevant features of the vibrational states of the system.

Three scarring POs, referred to as PO-A, PO-B, and PO-C, are involved in the frontier of scars, which is constituted by a series of broad ACs. The first two (PO-A and PO-B) are associated with the upper eigenstates in the series of ACs, while the third (PO-C) is associated with the lower eigenstates. Moreover, within the whole energy range, the last two (PO-C and PO-B) are isolated unstable POs, while the first one (PO-A) changes from isolated unstable to stable PO as quantum number increases (i.e., energy and Planck constant decrease) throughout the series. When these POs are quantized, yielding the corresponding semiclassical energies, the cases PO-A and PO-C give throughout the series a good agreement with the energies of the upper and lower eigenstates, respectively. However, the case PO-B evolves, as quantum number increases in the series, from a energy value close to the energy of the lower eigenstate, until a energy value close to the energy of the upper eigenstate. Indeed, the energies of both eigenstates corresponding to each AC in the series can be obtained through the semiclassical quantization of an {\em isolated unstable} PO, namely, case PO-C for the lower eigenstate and cases PO-B or PO-A (depending on the quantum number) for the upper eigenstate. And this is the main result of our work.

Additionally, we have found an approximate linear correlation in the frontier of scars that relates quantum number and Planck constant value at which each AC takes place. Extending the discrete domain of the quantum number in this relationship to a continuous domain, and applying the ``quantization'' condition, we have obtained the continuous semiclassical frontier between order and chaos, which matches the quantum frontier at (positive even) integer quantum numbers. Moreover, although the relationship is approximate, we can assume that the qualitative behavior of the quantized action derived from it is correct. Namely, as quantum number increases, the quantized action monotonically decreases from a vertical asymptote towards a horizontal asymptote. Assuming a positive deviation from the linear behavior as quantum number increases, we have conjectured values $n=6$ and $S\approx22$ a.u. for the vertical and horizontal asymtote, respectively (rather than values $n_0=4$ and $S_\infty\approx19$ a.u. obtained from the linear fitting), which are related to the first possible AC in the frontier of scars observed in the correlation diagram and the bifurcation where PO-C becomes an isolated unstable PO.

On the other hand, in order to calculate the non-trivial Maslov index of case PO-C, which is necessary for the semiclassical quantization, we have introduced a novel straightforward  method based on Lagrangian descriptors~\cite{Mancho.LDs.different_definitions}, this been the second relevant contribution of this paper. Notice that, in the cases PO-B and PO-A, the Maslov index was trivially obtained by counting the number of turning points in each degree of freedom. Eckhardt and Wintgen~\cite{Maslov.winding.number} proved that the Maslov index of a PO can be obtained by calculating the winding number of the invariant manifolds over one period, albeit the direct calculation of this parameter is mathematical demanding. However, we have shown how this winding number can be obtained by means of the easily calculation and depiction of the Lagrangian descriptors on a suitable surface of section along the corresponding PO.

\begin{acknowledgments}
This research was supported by the Ministry of Science and Innovation-Spain under Grant No.\ PID2021-122711NB-C21 ({\sc ChaSisCOMA} project).
\end{acknowledgments}

\end{document}